\documentclass[aps,prl,showpacs,twocolumn,amsmath,amssymb,superscriptaddress,secnumarabic]{revtex4-1}
\usepackage{graphicx}
\usepackage{hyperref}
\def\TN{\ensuremath{T_\text{N}}}
\def\TC{\ensuremath{T_\text{C}}}
\def\PFNO{\ensuremath{\text{Pb}(\text{Fe}_{0.5}\text{Nb}_{0.5})\text{O}_3}}
\begin{document}

\title{High-resolution structure studies and magnetoelectric coupling
of relaxor multiferroic Pb(Fe$_{0.5}$Nb$_{0.5}$)O$_3$}

\author{Hasung Sim}
\email{hssim@snu.ac.kr}
\affiliation{Center for Correlated Electron Systems, Institute for Basic Science (IBS), Seoul 151-747, Korea}
\affiliation{Department of Physics and Astronomy, Seoul National University, Seoul 151-747, Korea}

\author{Darren C.\ Peets}
\affiliation{Center for Correlated Electron Systems, Institute for Basic Science (IBS), Seoul 151-747, Korea}
\affiliation{Department of Physics and Astronomy, Seoul National University, Seoul 151-747, Korea}

\author{Sanghyun Lee}
\affiliation{Center for Correlated Electron Systems, Institute for Basic Science (IBS), Seoul 151-747, Korea}
\affiliation{Institute of Materials Structure Science and J-PARC Center, KEK, Tsukuba 305-0801, Japan}

\author{Seongsu Lee}
\affiliation{Neutron Science Division, Korea Atomic Energy Research Institute, Daejeon 305-353, Korea}

\author{T.\ Kamiyama}
\author{K.\ Ideda}
\author{T.\ Otomo}
\affiliation{Institute of Materials Structure Science and J-PARC Center, KEK, Tsukuba 305-0801, Japan}

\author{S.-{W}. Cheong}
\affiliation{Rutgers Center for Emergent Materials and Department of Physics and Astronomy, Rutgers University, Piscataway, New Jersey 08854, USA}

\author{Je-Geun Park}
\email{jgpark10@snu.ac.kr}
\affiliation{Center for Correlated Electron Systems, Institute for Basic Science (IBS), Seoul 151-747, Korea}
\affiliation{Department of Physics and Astronomy, Seoul National University, Seoul 151-747, Korea}


\begin{abstract}
\PFNO\ (PFN), one of the few relaxor multiferroic systems, has a
$G$-type antiferromagnetic transition at \TN~=~143~K and a
ferroelectric transition at \TC~=~385~K. By using high-resolution
neutron-diffraction experiments and a total scattering technique, we
paint a comprehensive picture of the long- and short-range structures
of PFN: (i) a clear sign of short-range structural correlation above
\TC, (ii) no sign of the negative thermal expansion behavior reported
in a previous study, and (iii) clearest evidence thus far of
magnetoelectric coupling below \TN. We conclude that at the heart of
the unusual relaxor multiferroic behavior lies the disorder between
Fe$^{3+}$ and Nb$^{5+}$ atoms. We argue that this disorder gives rise
to short-range structural correlations arising from O disorder in
addition to Pb displacement.
\end{abstract}

\maketitle

\section{Introduction}
It is rare in nature to find a single system which hosts two or more
ordered ground states out of otherwise unconnected degrees of
freedom. If multiple ordered ground states exist, a natural question
to ask is whether they are connected to one another. Although the same
electrons may be responsible for multiple degrees of freedom, allowing
coupling of the various forms of order, this is uncommon, making
materials in which they are coupled particularly
interesting. Magnetoelectric multiferroic materials, in which
ferroelectricity and magnetism coexist and are coupled, are an example
of just such a material.  The ability to control one form of order via
another offers a huge potential for technological applications and, at
the same time, poses new challenges for our understanding of how
distinct degrees of freedom as disparate as bulk polarization and
magnetization can be coupled to one
another~\cite{Cheong2007,Fiebig2005}.  Here the sought-after coupling,
so-called magnetoelectric effects, can lead to better manipulation of
unusual multiferroic behavior and to exotic excitations of mixed
character. Despite the huge interest, however, the origin of
magnetoelectric coupling has often proven challenging to address
experimentally for a given material.

Lead iron niobate \PFNO\ (PFN) is a multiferroic material with a
ferroelectric transition at \TC~=~385~K~\cite{Kania2009} and an
antiferromagnetic transition at
\TN~=~143~K~\cite{Kania2009,Mishra2010,Laguta2013,Wang2004} which is
known to be $G$-type~\cite{Ivanov2000}. It is noteworthy for its high
dielectric constant, which changes at both the
ferroelectric~\cite{Mishra2010,Sitalo2009,Fraygola2011} and the
magnetic transitions~\cite{Kania2009,Fraygola2011,Lente2008} and is
strongly frequency dependent.  Its reported high dielectric constant
makes it a suitable candidate material for multilayer ceramic
capacitors among other electronic devices. The reported strong
frequency dependence, making it a rare example of a relaxor
multiferroic, most likely arises from disorder at the magnetic $B$
site of the perovskite structure. Since many ferroelectric systems
have disordered magnetic ions on the $B$ site, a full understanding of
PFN might well have wider implications for finding or optimizing other
relaxor multiferroic materials for potential applications.

Despite the interest in the physical properties of PFN, however,
several features of the underlying crystal structure still remain
unresolved. For example, two competing space groups,
$R3m$~\cite{Ivanov2000} and $Cm$~\cite{Singh2007,Lampis1999}, have
been proposed for the low-temperature structure, and there is an
unconfirmed report of negative thermal expansion below the
antiferromagnetic transition~\cite{Singh2007}. In attempting to
characterize the relaxor behavior, several groups have investigated
the local structure~\cite{Mesquita2012,Jeong2011}, which is closely
associated with relaxor behavior in ferroelectrics, and discussed the
possibility of structural disorder both
experimentally~\cite{Jeong2011,Kolesova1993} and
theoretically~\cite{Wang2001}.  These studies notwithstanding, the key
questions, we believe, are yet to be fully answered. For instance, it
will be interesting to know how the relaxor behavior is connected to
the short-range local structure. More importantly, one first has to
know the exact details of the local distortion and, if possible, the
structural basis of the sought-after magnetoelectric coupling.

In order to address these questions, we have undertaken full
high-resolution neutron powder-diffraction studies as well as total
scattering experiments using two state-of-the-art instruments: S-HRPD
and NOVA, both at J-PARC, in Tokai, Japan. For this kind of study it
is essential to be able to examine both long- and short-range
structures simultaneously as performed for other ferroelectric
materials~\cite{Jeong2011b}. By combining data obtained from both
instruments, we conclude that the low-temperature space group is $Cm$
with clear signs of the short-range structure surviving even above the
ferroelectric transition temperature, offering an explanation for the
relaxor behavior. Neither experiment showed signs of negative thermal
expansion. We then uncovered structural signatures of the
magnetoelectric coupling by analyzing the temperature dependence of
the electric polarization calculated from the structure
parameters. Our detailed analysis of the local structure leads us to
the conclusion that both Pb displacement and O disorder are exhibited
in this material.

\section{Experimental Details}
About 7~g of polycrystalline PFN samples were prepared by a standard
solid-state method. Stoichiometric PbO, Fe$_2$O$_3$, and Nb$_2$O$_5$
were mixed, then calcined in air at 850\,$^\circ$C for 2~days. After
calcination the products were ground, pressed into pellets, and then
sintered at 1050\,$^\circ$C for 1~day. The samples were verified to be
single phase with a Rigaku Miniflex II x-ray diffractometer as well as
the later high-resolution neutron diffraction studies which form the
main body of this paper.

High-resolution neutron time-of-flight powder-diffraction experiments
were carried out using the S-HRPD beamline at J-PARC, Tokai, Japan, on
a powder sample in a cylindrical vanadium can, at ten temperatures
from 10 to 300~K. Rietveld refinement of the diffraction data was
performed using {\footnotesize
  FULLPROF}~\cite{Rodriguez-Carvajal1993}. The subsequent total
scattering experiments were carried out using a vanadium-nickel sample
can at the NOVA beamline at J-PARC with a maximum $Q$ value of
$Q_{max} = 30$~\AA$^{-1}$ at five temperatures between 62 and
453~K. Sample environment constraints prevented measurements above the
ferroelectric transition temperature at S-HRPD, so NOVA was also used
to collect diffraction patterns at selected higher temperatures. For
the local structure study, a radial distribution function was
calculated from the total scattering data and analyzed using
{\footnotesize PDFGUI}~\cite{Farrow2007}. The experimentally obtained
$S(Q)$ was then Fourier transformed into real space to obtain the
atomic pair distribution function $G(r)$ as follows:
\begin{displaymath}
G(r) = 4\pi r\left[\rho(r) - \rho_0\right] = 
\frac{2}{\pi}\int^{Q_{max}}_0\left[S(Q)-1\right]\sin Qr\,dQ \text{,}
\end{displaymath}
where $\rho(r)$ and $\rho_0$ are the atomic number density and the
average number density, respectively, and $Q_{max}$ is the maximum
scattering wave vector. $G(r)$ was modeled using $4 \times 4 \times 4$
supercells for several different models of the short-range structure.

\section{Results and discussion}

\begin{figure}[htb]
\includegraphics[width=0.85\columnwidth]{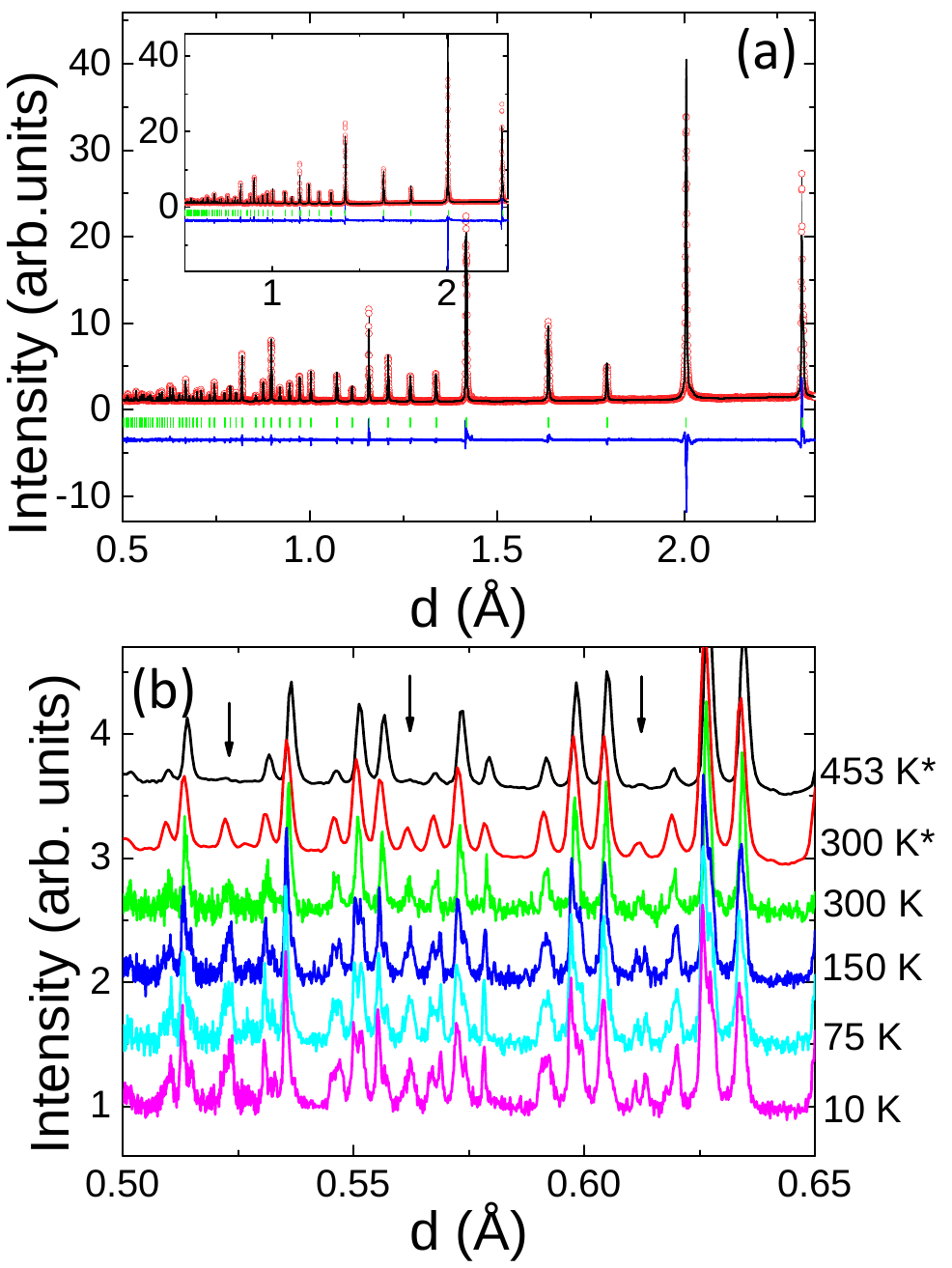}
\caption{\label{fig1}(a) Neutron powder-diffraction patterns with
  refined results in the $Cm$ and $R3m$ (inset) space groups,
  respectively, at 300~K. Circles represent data, a line the
  refinement, and the line at the bottom the residual, whereas green
  vertical bars mark the Bragg positions. (b) Diffraction patterns at
  several temperatures. The upper two datasets (starred) are NOVA
  results, and the others are S-HRPD results. Several peaks disappear
  above \TC~=~385~K due to the structural transition (arrows).}
\end{figure}

The low-temperature space group of PFN is thought to be the monoclinic
space group $Cm$, but as discussed, the $R3m$ space group has also
been proposed as an alternative structure; to the best of our
knowledge it is still not settled which one of the two space groups is
correct. Figure~\ref{fig1}(a) shows the neutron powder-diffraction
pattern and its refinement results in the $Cm$ and $R3m$ (inset) space
groups at 300~K. We assumed that Fe and Nb are randomly distributed at
the same crystallographic
position~\cite{Ivanov2000,Singh2007,Lampis1999} and allowed this
sublattice to shift relative to Pb but did not split any sites at this
stage. Satisfactory refinements have previously been reported in both
$R3m$~\cite{Ivanov2000} and $Cm$~\cite{Singh2007}; we rely on an
additional approach to distinguish these space groups. The data are
well explained by both structures, although the quality of the
refinement is slightly better in $Cm$, but this is not surprising
since the $Cm$ structure model has more free
parameters. Table~\ref{table1} summarizes the refined structural
parameters for the 300~K data in the $Cm$ and $R3m$ structures.
Results are described in the rhombohedral setting of the $R3m$
structure for comparison with Ref.~\onlinecite{Ivanov2000}. One can
also convert the lattice parameters of the $Cm$ structure to
pseudocubic notation to ease comparison with the paraelectric
structure as performed in Fig.~\ref{fig3}(a)~\footnote{Multiplying by
  the matrix
$\left(
\begin{smallmatrix}
1/\sqrt{2} & 1/\sqrt{2} & 0\\
-1/\sqrt{2} & 1/\sqrt{2} & 0\\
0 & 0 & 1
\end{smallmatrix}
\right)$ 
converts the monoclinic structure to the pseudocubic structure.}.

\begin{table}[htb]
\caption{\label{table1}Refined structure parameters at 300~K in the
  $Cm$ (with $b$ as the unique monoclinic axis and cell choice 2) and
  $R3m$ structures.  The unsplit Pb position is taken as the origin in
  $Cm$, whereas the Fe/Nb position is set to (0.5,0.5,0.5) for $R3m$
  to aid comparisons to the literature.}
\begin{center}
\begin{tabular}{llllr}
\hline \hline
{\it Cm} & \multicolumn{4}{p{6.8cm}}{$a_{m}=5.6819(1)$~\AA, $b_{m}=5.6738(1)$~\AA, $c_{m}=4.01202(5)$~\AA, $\beta=89.896(2)^{\circ}$}\\
Ions & \multicolumn{1}{c}{$x$} & \multicolumn{1}{c}{$y$} & \multicolumn{1}{c}{$z$} & $B$ (\AA$^2$)\\
\hline
Pb & 0.0000 & 0.0000 & 0.0000 & 1.82(8) \\
Fe/Nb & 0.4669(9) & 0.0000 & 0.5165(18) & 0.34(3) \\
O1 & 0.4333(13) & 0.0000 & 0.0002(26) & 0.54(3) \\
O2 & 0.2129(20) & 0.2483(7) & 0.4989(29) & 0.54(3) \\
\multicolumn{5}{l}{$R_p$: 5.26, $R_{wp}$: 7.68, $R_{exp}$: 4.13, $\chi^2$: 3.45} \vspace{6pt}\\
$R3m$ & \multicolumn{4}{l}{$a_{r}=4.01389(3)$~\AA, $\alpha = \beta = \gamma =89.9223(2)^\circ$}\\
\hline
Pb & 0.0205(6) & 0.0205(6) & 0.0205(6) & 2.45(5) \\
Fe/Nb & 0.5000 & 0.5000 & 0.500 & 0.27(2) \\
O & 0.4772(2) & 0.4772(2) & -0.0092(5) & 0.58(2) \\
\multicolumn{5}{l}{$R_p$: 5.48, $R_{wp}$: 8.01, $R_{exp}$: 4.13, $\chi^2$: 3.76} \\
\hline \hline
\end{tabular}
\end{center}
\end{table}

In the temperature-dependent data, magnetic peaks can be observed at
$d = 1.55$ and 1.85~\AA, consistent with the reported $G$-type
antiferromagnetic structure~\cite{Ivanov2000}. Several structural
peaks, shown in Fig.~\ref{fig1}(b), are absent above the ferroelectric
transition of 385~K as seen comparing the data taken at 453 and 300~K;
these peaks are marked by arrows. We note that we can refine the
highest temperature data in the paraelectric phase using the
previously reported space group
$Pm\overline{3}m$~\cite{Lampis1999,Kolesova1993,Kolesova1999}; however,
as we will discuss later, there are clear signs of short-range
structure in the total scattering data even at 453~K.

\begin{figure}
\includegraphics[width=0.85\columnwidth]{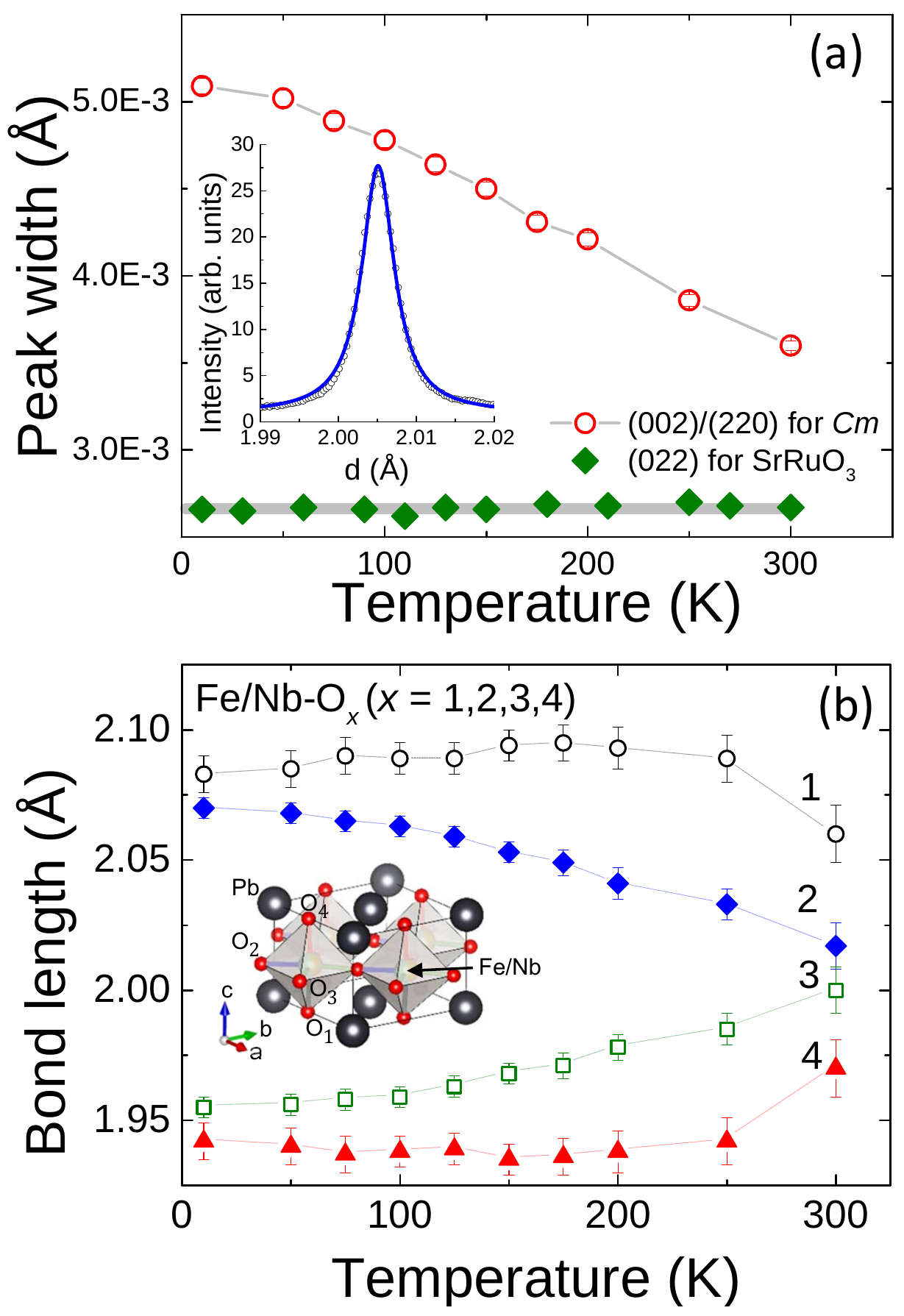}
\caption{\label{fig2}(a) Temperature dependence of the peak width of
  the (002)/(220) Bragg peak in the $Cm$ structure with the raw data
  at 10~K shown in the inset. Error bars are smaller than the symbol
  size. The data for the (022) peak of SrRuO$_3$ were taken after
  Ref.~\cite{Lee2013} (see the text). (b) Temperature dependence of
  Fe/Nb--O bond lengths.  The inset depicts the structure of PFN with
  the $Cm$ space group: There are four different Fe/Nb--O bonds
  denoted by the numbers.}
\end{figure}

In order to distinguish the two candidate low-temperature space
groups, the temperature dependence of the peak widths for several
nuclear Bragg peaks as obtained from fits to a Lorentzian line shape
was examined, and a search was performed for peaks that are split in
one space group but not the other. The large number of reflections in
this low-symmetry structure makes it difficult to find nonoverlapping
peaks, but several good candidates were located. The (002) and (220)
Bragg reflections of the $Cm$ structure, shown in the inset of
Fig.~\ref{fig2}(a), were found to be the most suitable: In the $R3m$
space group the resulting peak contains only one unique reflection,
(200). No clear splitting is observed at low temperatures, but the
peak width increases markedly on cooling as shown in
Fig.~\ref{fig2}(a), which is exactly opposite to what one expects from
any conventional thermally activated broadening process.  Similar
temperature dependence is observed in other peaks, which are split in
$Cm$ and not in $R3m$. As we cannot find a peak from a well-separated
Bragg reflection in our data because of the low symmetry, we choose
SrRuO$_3$ measured under almost identical conditions in order to
demonstrate how the width of a single Bragg peak behaves as a function
of temperature when measured at the S-HRPD beamline. In
Fig.~\ref{fig2}(a), we plotted the (022) peak of
SrRuO$_3$~\cite{Lee2013} where the width of the single Bragg peak
remains temperature independent over the measured temperature range as
expected. The unusual peak broadening seen in the (002)/(220) peak of
PFN is naturally explained by a splitting of two reflections, which
strengthens on cooling, and would appear to exclude the $R3m$
structure.  The origin of this unusual temperature dependence can be
found by examining the Fe/Nb--O bond lengths and their temperature
dependence, shown in Fig.~\ref{fig2}(b). Two aspects are particularly
noteworthy: First, there are already four different Fe/Nb--O bond
lengths even at room temperature [see the inset in Fig.~\ref{fig2}(b)
  for the four oxygen atoms with different Fe/Nb--O bond lengths
  denoted by different numbers]; second, the splitting of the Fe/Nb--O
bond lengths only continues to grow upon cooling. Thus, the larger
distortion of the (Fe/Nb)O$_6$ octahedron lies at the heart of the
unusual peak broadening seen in Fig.~\ref{fig2}(a), favoring the $Cm$
space group.

\begin{figure*}[htb]
\begin{center}
\includegraphics[width=0.75\textwidth]{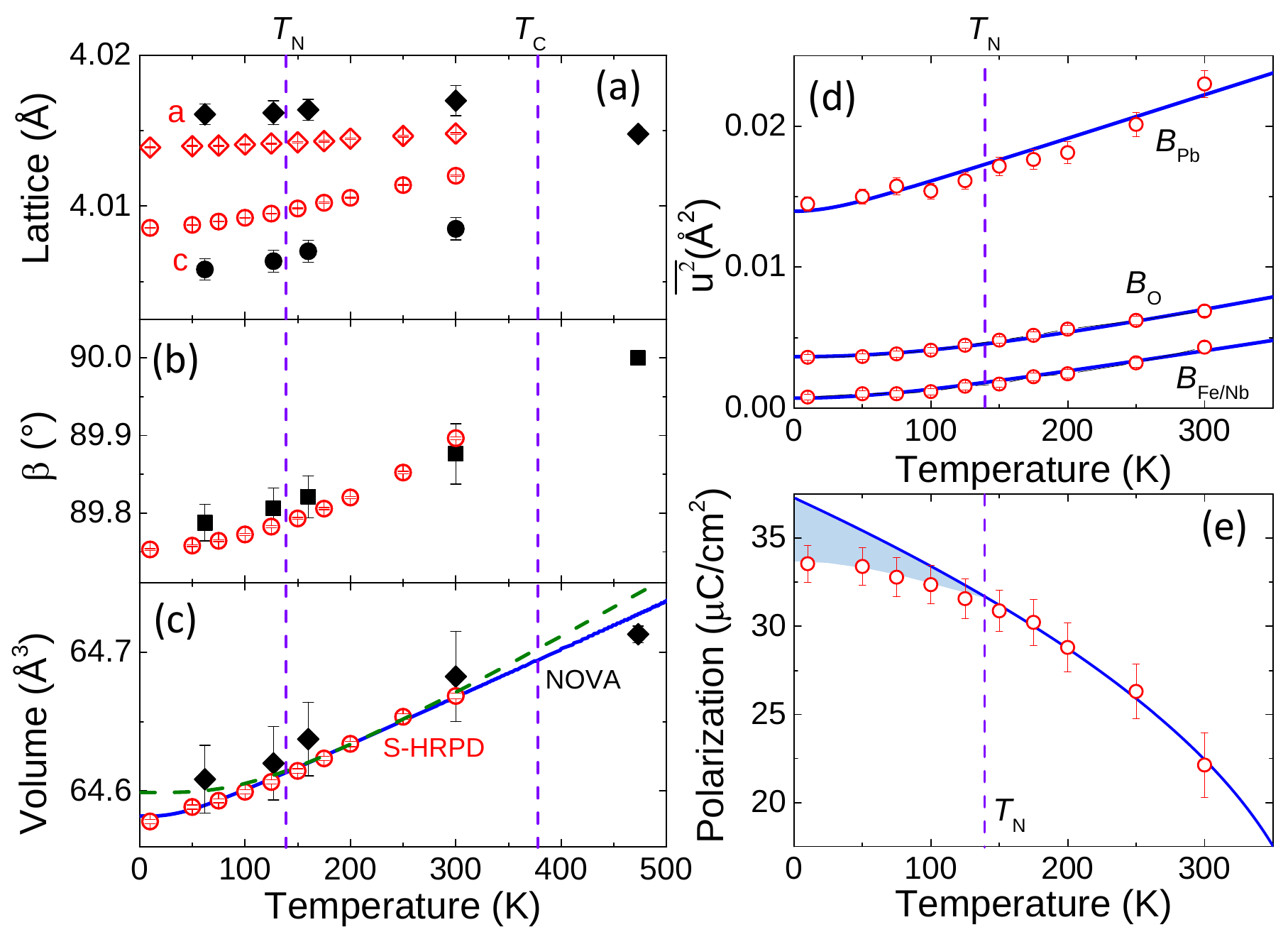}
\caption{\label{fig3}Temperature dependence of (a) lattice parameters,
  (b) monoclinic angle, (c) unit-cell volume, (d) thermal parameter,
  and (e) calculated polarization. The refined unit-cell volume
  (symbols) is shown together with the theoretical curve obtained from
  the Debye-Gr\"uneisen formula, and polarization was calculated from
  the refined atomic positions.}
\end{center}
\end{figure*}

The temperature dependence of the lattice constants and unit-cell
volume are summarized in Fig.~\ref{fig3}. We also plot the monoclinic
angle in pseudocubic notation. Square symbols represent NOVA results,
whereas red circles indicate those from S-HRPD. Refinements were
performed in the $Pm\overline{3}m$ structure above \TC\ and the $Cm$
structure below \TC. Both the lattice constants and the unit-cell
volume follow a consistent trend without any clear anomaly at the
antiferromagnetic transition.  Regarding the previous claim of
negative thermal expansion~\cite{Singh2007}, we note that there is no
evidence of such an effect in the data presented here, taken using
either instrument, although it is also important to note that lattice
parameters from x-ray diffraction are normally more precise. It is an
open question to us why our data collected from two different
instruments do not show the negative thermal expansion claimed in
Ref.~\onlinecite{Singh2007}.

In order to further analyze the temperature dependence of the
unit-cell volume, the Debye-Gr\"uneisen formula~\cite{Wood2002} was
used
\begin{align}
V(T) &= \frac{V_{0}U(T)}{Q-bU(T)} + V_0\text{,}\\
U(T) &= 9Nk_{B}T\left(\frac{T}{\Theta_{D}}\right)^3 \int^{\frac{\Theta_{D}}{T}}_{0}{\frac{x^3}{e^x-1} dx},
\end{align}
where $U(T)$ is total internal energy, $Q=V_0B_0/\gamma$,
$b=(B'_0-1)/2$, $\gamma$ is the Gr\"uneisen parameter, $B_0$ and
$B'_0$ are the bulk modulus and its first derivative with pressure,
$N$ represents the number of atoms in the unit cell, $k_B$ is the
Boltzmann constant, and $V_0$ is volume at zero temperature.  Fitting
(shown by a solid line) results in a Debye temperature $\Theta_D$ of
only 150~K, whereas $b=1.5$, $V_0=64.582$~\AA$^3$, and
$Q=3.87(6)\times 10^{-17}$~J are similar to parameters used for other
materials~\cite{Wu1983,Lee2006,Park2010}. We note that the resulting
Debye temperature is abnormally small, leading to a bulk modulus
around 600~GPa for a typical Gr\"uneisen parameter of order one, which
is high for an oxide: For example, we earlier found bulk moduli of
120~GPa for YMnO$_3$~\cite{Kozlenko2005} while 250~GPa has been
reported for MgSiO$_3$~\cite{Andrault2001}. Note that this assumption
about the Gr\"uneisen parameter is crude, and so our estimate of the
bulk modulus should be taken with caution. Moreover a true test of the
bulk modulus needs to be performed by measuring the volume measurement
under pressure. On the other hand, fixing the Debye temperature to the
more typical 430~K led to a noticeably poorer fit (dashed line) with
$b=1.5$, $V_0=64.595$~\AA$^3$, and $Q=2.9(1)\times 10^{-17}$~J. This
could be consistent with an anomaly below \TN\ but opposite in sign to
that reported previously.

Figure~\ref{fig3}(d) shows the temperature dependence of atomic
displacement parameters $\overline{u^2}$ for each atom. The thermal
parameter of Pb is much larger than those of O and Fe/Nb, almost
comparable to those of other Pb-containing ferroelectric
materials~\cite{Lampis1999} and consistent with a previous
report~\cite{Jeong2011}. The temperature dependence was modeled using
\begin{align}
\overline{u^2} &= \left(\frac{145.55T}{M{\Theta_D^i}^2}\right)
\varphi\left(\frac{\Theta_D^i}{T}\right)+A^i\\
\varphi\left(\frac{\Theta_D^i}{T}\right) &= \frac{T}{\Theta_D^i}
\int^\frac{T}{\Theta_D^i}_0 \frac{x}{e^x-1}dx\text{,}
\end{align}
where $i$ represents each atomic species (Pb, Fe/Nb, O), and $A^i =
36.39/M^i\Theta_D^i$ is related to zero-point energy of the atoms
concerned with atomic mass $M^i$\cite{Wood2002}. However, it was
necessary to add constant offsets to $A$ to achieve better
agreement. The final result, denoted by lines in Fig.~\ref{fig3}(d),
indicated effective Debye temperatures of 150~K for Pb, 680~K for O,
and 370~K for Fe/Nb.

Using the structural information, we then searched for possible
experimental evidence of magnetoelectric coupling.  PFN is known to
exhibit a frequency-dependent anomaly in the dielectric constant at
\TN, but little else is known of the magnetoelectric coupling. From
the high-accuracy structural information presented here, it is
possible to calculate the expected ferroelectric polarization,
starting from the centrosymmetric $Pm\overline{3}m$ space group. Using
a similar approach, we were previously able to show a negative
magnetoelectric coupling in one of the best-studied multiferroic
compounds, BiFeO$_3$~\cite{Lee2013b}. For simplicity, the nominal
valences of the atoms were used as a starting point: Pb$^{2+}$,
Fe$^{3+}$, Nb$^{5+}$, and O$^{2-}$. Since the actual valences,
so-called Born effective charges, which include dynamic terms, might
well differ from the nominal charge values~\cite{King-Smith1993}, the
discussion below may not be fully quantitative, but it should be
qualitatively correct. Using the above assumptions, the electric
polarization was calculated with respect to the paraelectric
$Pm\overline{3}m$ phase. As shown in Fig.~\ref{fig3}(e), the estimated
polarization at room temperature is around 20\,$\mu$C/cm$^2$, which is
twice that reported~\cite{Kania2009,Mishra2010} in polycrystalline
samples. Considering the rough assumptions made and the fact that the
experiments were performed on powder samples, we consider this degree
of deviation acceptable.  The temperature dependence was then
parametrized using Ginzburg-Landau analysis, assuming that the
electric polarization follows the usual first-order temperature
dependence: $F(P) = \frac{\alpha_2}{2} P^2 + \frac{\alpha_4}{4} P^4 + \frac{\alpha_6}{6} P^6$,
where $F$ is the Ginzburg-Landau free energy and $P$ is the
polarization.  A fit to all points produced an unphysically high Curie
temperature, whereas constraining the Curie temperature led to clear
systematic trends in the residuals; a fit to only the points above the
N\'eel transition, shown in Fig.~\ref{fig3}(e), predicted a Curie
temperature of 380~K in close agreement with experiment, but the data
appear to deviate from this fit below \TN. This discrepancy, marked by
shading in the figure, is the clearest experimental evidence yet of
the magnetoelectric coupling in PFN.

\begin{figure*}
\begin{center}
\includegraphics[width=0.85\textwidth]{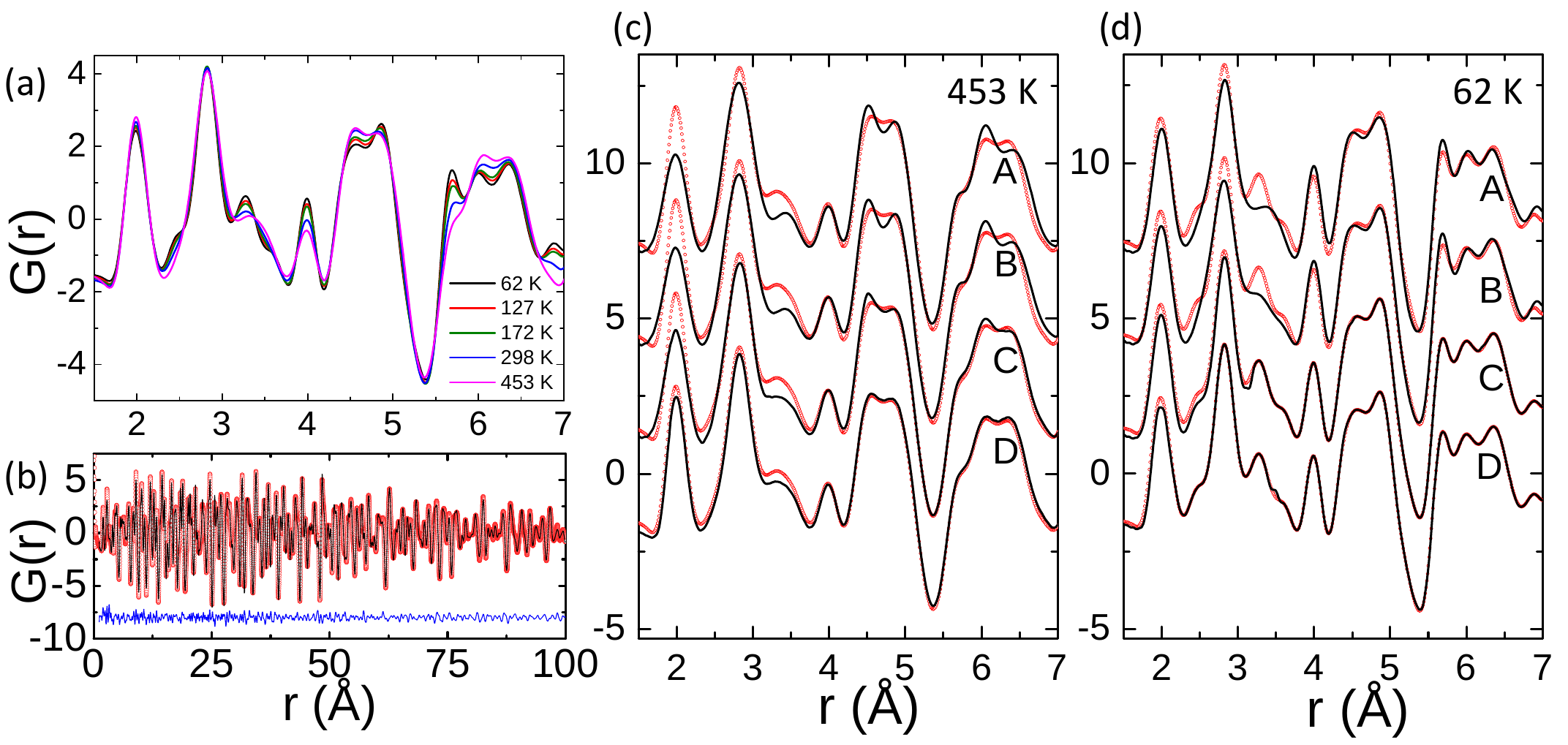}
\caption{\label{fig4}(a) Radial distribution function $G(r)$ at
  several temperatures, calculated from NOVA results. Gradual changes
  may be observed with temperature at several regions of interest (see
  the text). (b) Radial distribution function $G(r)$ is compared with
  fitting results over a wide length-scale range for the 62~K
  data. The symbols represent the data points, whereas the line is the
  fitting results with the difference curve shown at the bottom. The
  local structure is compared with several models at (c) 453~K and (d)
  62~K. Circles are calculated from experimental data, and lines are
  fits to several models: Model A assumes no local structure; model B
  assumes Pb displacements; model C considers Pb disorder; and model D
  combines Pb displacements with O disorder. Traces have been offset
  vertically for clarity.}
\end{center}
\end{figure*}

For further analysis of the short-range structure, Fig.~\ref{fig4}(a)
shows how the radial distribution function $G(r)$ calculated from the
NOVA data changes gradually with temperature. Of note are the several
regions where peaks' intensities increase on cooling. At these short
length scales, corresponding to large momentum transfers $Q$, we can
ignore contributions from magnetic scattering because they are usually
small compared to nuclear scattering~\cite{billinge}. As shown in
Fig.~\ref{fig4}(b), the long-range structure up to 100~\AA\ can be well
explained by the $Cm$ space group.

First, we consider the short-range structure at high temperatures
above the ferroelectric transition. Using the data taken at 453~K, we
tested four different models to explain the data and clarify the local
structure, all variants of the $Cm$ structure, and all explained in
more detail below: (A) one without local short-range structure, (B)
one with Pb displacements as considered previously~\cite{Jeong2011},
(C) one with Pb disorder, and (D) one with both Pb displacement and O
disorder. For all models, Fe and Nb are treated as an average
pseudoatom as in the refinement. Models B--D take model A as their
starting point and allow additional freedom in the atomic
positions. In departing from model A, $C$ centering was not
retained. As shown in Fig.~\ref{fig4}(c), the agreement is indeed
improved by including Pb displacement and O disorder in our model
calculation. Upon closer inspection, model A is found not to
satisfactorily explain the intensities of peaks at $r = 2$ and
3.5~\AA, and agreement with experimental results is poor at longer
distances. It is particularly surprising that, even in the nominally
paraelectric phase well above the ferroelectric transition, there is
clear evidence of short-range structure. We note that similar behavior
is often found in relaxor ferroelectric
materials~\cite{Jeong2005,Egami2003}. Thus, our result adds further
weight to the previous claim that PFN is a rare relaxor multiferroic.

Having concluded that the simplistic model A does not adequately
explain even the high-temperature local structure, we now consider
other models which might describe the data at all temperatures. Given
lead's large thermal parameter even after accounting for the offset
between the Pb and Fe/Nb sublattices in the $Cm$ structure (model A)
as shown in Fig.~\ref{fig3}(d), it is natural to think of further Pb
displacements as an origin of short-range structure as is often the
case in closely related Pb-containing perovskite
systems~\cite{Lampis1999,Malibert1997} and as considered
previously~\cite{Jeong2011}. At 62~K, model B requires that the lead
atoms shift by ($-$0.019,0.03,0.012) in fractions of the pseudocubic
unit cell, less than previously reported~\cite{Jeong2011}. Moving the
Pb atoms introduces several peaks between $r = 3$ and 4~\AA, but their
intensities are much smaller than in the experimental results, and the
peak at $r = 2.5$~\AA\ is not reproduced.  Above the ferroelectric
transition, symmetry constrains the Pb displacement to be
$(\eta,\eta,\eta)$, where $\eta = 0.02$ at 453~K. Displacing the Pb
atoms in concert, which preserves the unit cell, is quite a simplistic
model, and allowing the Pb atoms more freedom of movement should
better model the real system.  Accordingly, model C was introduced
with the possibility of random shifts in the Pb positions.

Model C simulates the effect of Pb disorder using a $4\times 4\times
4$ supercell in which each lead atom was allowed to move freely from
its original position. Although this model was able to better model
the experimental radial distribution function, at least at low
temperatures, the average Pb displacement from its already displaced
$Cm$ position is 0.2~\AA\ in this model. This is very large compared to
Pb's atomic thermal motion in Fig.~\ref{fig3}(d), approaches the upper
limit observed in the strongest Pb-based ferroelectrics and BiFeO$_3$
with one of the highest polarization values of $\sim
86\,\mu$C/cm$^2$~\cite{Lee2013b}, and would suggest an electric
polarization far exceeding that
reported~\cite{Kania2009,Mishra2010}. The comparative success of model
C relative to model B is most likely a result of the significantly
higher number of free parameters. Although model C is ruled out on the
basis of the extremely large displacements compared to the reported
polarization values, the fact that it required unphysical
displacements of Pb implies that other atoms must also be
considered. It has already been shown that the Fe/Nb atoms do not
exhibit any significant displacement from their ideal
positions~\cite{Jeong2011}. That leaves oxygen, which has a larger
thermal parameter and which is well known to depart from its ideal
position in most perovskite structures, including in several
Pb-containing phases which exhibit oxygen
displacement~\cite{Dmowski2000,Baldinozzi1992,Chen2000} or the random
rotation of oxygen polyhedra~\cite{Egami2003}. Model D allows the
oxygen atoms to shift.

A further justification for considering O disorder comes from the
reported random distribution of neighboring Fe$^{3+}$ and
Nb$^{5+}$~\cite{Ivanov2000} and the lack of any features in our data
that would suggest even short-range cation order. The magnetic moments
calculated from the long-range refinement are 3.4(6)\,$\mu_B$ at 62~K
(NOVA) and 3.3(3)\,$\mu_B$ at 10~K (S-HRPD), corresponding to the high
spin state of Fe$^{3+}$, which would make the two cations' sizes
essentially identical and eliminate the common source of disorder. The
next most probable reason for oxygen atoms to depart from their ideal
positions is the displacements of the Pb$^{2+}$ cations from their
already-shifted positions, which would be expected to cause small
rotations of the (Fe,Nb)O$_6$ octahedra. This would be frustrated,
unless the Pb displacements ordered in one of a few very specific ways
which would be expected to produce additional nuclear reflections. An
orthogonal possibility arises from the electric-field gradients
created by having a random distribution of Fe$^{3+}$ and Nb$^{5+}$
cations. An oxygen atom situated between an iron and a niobium site
feels an electric-field gradient that will polarize it. Meanwhile,
Nb$^{5+}$, having the electronic configuration of krypton, will be
unable to share electron density with oxygen, which $3d^5$ Fe$^{3+}$
can, using its weakly antibonding $e_g^*$ orbitals. In fact, iron,
being a $3d$ transition metal, should have significant on-site
electron repulsion and may benefit by donating some of its electron
density to its oxygen ligands. All of these electronic effects would
produce a net flow of electron density away from iron sites and toward
niobium sites with many oxygen atoms caught in the middle. In model D,
we combined lead displacement (model B) with oxygen disorder using a
$4\times 4\times 4$ supercell. Here we allowed oxygen atoms to move
freely while all lead atoms were constrained to move together. As one
can see in Figs.~\ref{fig4}(c) and \ref{fig4}(d), this model
apparently fits the experimental results very well. The fitted atomic
displacement in fractions of the pseudocubic unit cell for lead is
(0.010,$-$0.004,0.008) at 62~K and ($-$0.0051) at 453~K, which is
reduced compared to model B. For oxygen, the average shifts relative
to the original site are (0.0195) at 62~K and (0.0203) at 453~K in
fractional coordinates, which is around 0.08~\AA. As is clear in
Fig.~\ref{fig4}, of the models considered here, model D is the most
successful at describing the experimental results, and it does so in a
physically reasonable manner.

\begin{figure}[htb]
\includegraphics[width=\columnwidth]{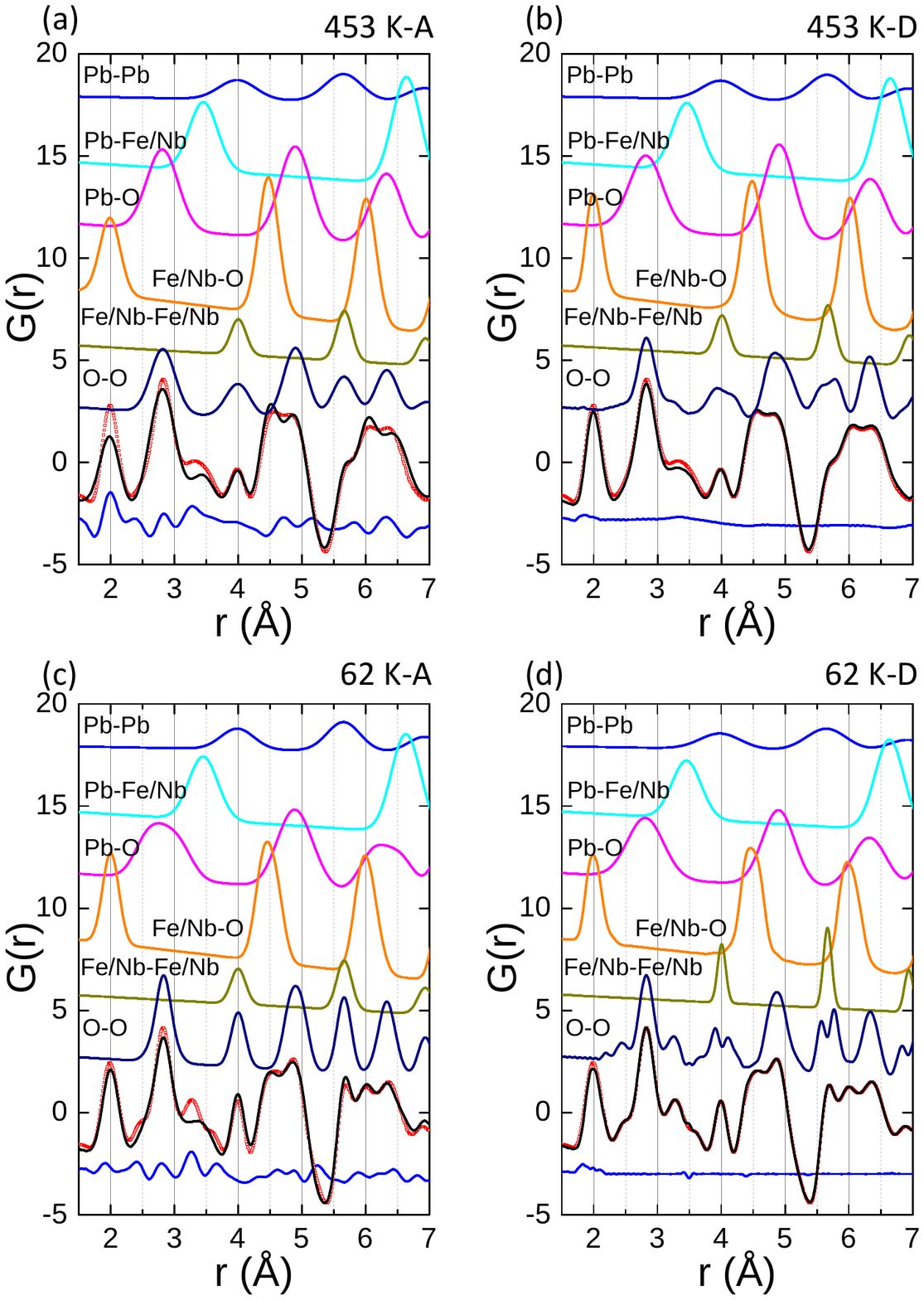}
\caption{\label{fig5} Atomic position correlations at 453(62)~K with
  (a) [(c)] no local structure and (b) [(d)] both Pb displacement and
  O disorder. The top six lines show atom-atom radial correlation
  functions as labeled, followed by the experimental results
  (circles), best fit (line), and finally its residuals (bottom
  line). Traces have been offset vertically for clarity.}
\end{figure}

In Fig.~\ref{fig5}, we deconvoluted the contributions of correlations
from each pair of atoms using models A and D for the
62 and 453~K data shown in Fig.~\ref{fig4}. In each panel, the
top six lines show correlation functions between individual
pairs of atoms as labeled. Below this are the experimental
results and best fit and finally the residuals. A comparison
of panels (a) with (b) in Fig.~\ref{fig5} reveals that O--O
correlations develop a great deal of additional structure at
both temperatures, explaining well the additional peaks at
$r = 2.5$--4~\AA.

\section{Conclusion}

In conclusion, we have undertaken both high-resolution
powder-diffraction and total scattering experiments on PFN in order to
shed light on the unusual relaxor multiferroicity in this material,
focusing on four key aspects of the physical properties: (i)
determination of the low-temperature structure, (ii) the experimental
examination of the reported negative thermal expansion behavior, (iii)
possible structural evidence of magnetoelectric coupling, and (iv) the
nature of the relaxor behavior. By combining both long- and
short-range diffraction data, it is possible to provide new insight on
each of the four issues. First, the high-resolution
neutron-diffraction data are more consistent with the $Cm$ space group
at low temperatures as reported based on synchrotron
data~\cite{Singh2007} and exclude the previously proposed
$R3m$~\cite{Ivanov2000} structure. However, there are no signs of the
negative thermal expansion reported in Ref.~\onlinecite{Singh2007}
with the data instead being well fit by the conventional
Debye-Gr\"uneisen formula. We succeeded in finding a structural
signature of the magnetoelectric coupling by Ginzburg-Landau analysis
of the calculated electric polarization. Finally, the data clearly
show that over the entire temperature range covered in this study
there exists a short-range structure in addition to the global
structure of $Pm\overline{3}m$ or $Cm$. This local structure occurs
through Pb displacement as well as O disorder. That this local
structure is present in PFN, even in the paraelectric phase, as seen
in relaxor ferroelectric materials may hold the key to the strong
frequency dependence seen in the dielectric constant of PFN.

\section{Acknowledgements}
We thank I.-K.\ Jeong for useful discussions. This
work was supported by IBS-R009-G1. The neutron scattering
experiment was approved by the Neutron Science
Proposal Review Committee of J-PARC/MLF (Proposal
No.\ 2012A0017) and supported by the Inter-University Research
Program on Neutron Scattering of IMSS, KEK.

\bibliography{PFN}
\end{document}